\documentclass[a4paper]{article}
\usepackage{amsmath}
\usepackage{amssymb,amsfonts,textcomp}
\usepackage[T1]{fontenc}
\usepackage[english]{babel}
\usepackage[utf8]{inputenc}
\usepackage{color}
\usepackage{array}
\usepackage{supertabular}
\usepackage{hhline}
\usepackage{hyperref}
\usepackage{changepage}
\usepackage{hanging}
\usepackage{setspace}
\usepackage[pdftex]{graphicx}

\newcommand\textstyleInternetlink[1]{\textcolor[rgb]{0.019607844,0.3882353,0.75686276}{#1}}
\makeatletter
\newcommand\arraybslash{\let\\\@arraycr}
\makeatother
\makeatother
\author{Microsoft-Konto}
\date{2023-12-21}

\begin{document}
\sloppy
\onehalfspacing 

\textbf{Influence of Li-stoichiometry on electrical and acoustic properties and temperature
stability of Li(Nb,Ta)O$_{3}$ solid solutions up to 900~°C}

\bigskip

\textit{Éva Tichy-Rács*, Stepan Hurskyy, Uliana Yakhnevych, Piotr Gaczy{\'n}ski, Steffen Ganschow, Holger Fritze and Yuriy Suhak}

\bigskip

{É. Tichy-Rács, S. Hurskyy, U. Yakhnevych, P. Gaczy}{\'n}{ski, H. Fritze and Y. Suhak}

Institute of Energy Research and Physical Technologies, University of Technology Clausthal,

Am Stollen
19B, 38640 Goslar, Germany

E-mail: \textstyleInternetlink{eva.tichy-racs@tu-clausthal.de}

\bigskip

S. Ganschow

Leibniz-Institut für Kristallzüchtung, Max-Born-Straße 2, 12489 Berlin, Germany

\bigskip


\bigskip

The current work is focused on the impact of the lithium stoichiometry on electrical conductivity, acoustic properties and high-temperature stability of single crystalline Li(Nb,Ta)O$_{3}$ at high temperatures. The crystals grown from Li-deficient melts were treated by the vapor transport equilibration (VTE) method, achieving near stoichiometric Li-content. It is shown, that the VTE-treated specimens generally exhibit lower conductivity at temperatures below 800~°C, which is attributed to the reduced number of Li-vacancies in near stoichiometric Li(Nb,Ta)O$_{3}$, provided that the Li-ion migration dominates the conductivity in this temperature range. Further, it is shown, that above 600--650~°C different mechanism increasingly contributes to the conductivity, which is consequently attributed to the electronic conduction. Further, it is shown that losses in LNT strongly increase above about 500~°C, which is interpreted to originate from conductivity-related relaxation mechanism. Finally, the thermal stability of Li(Nb,Ta)O$_{3}$ is evaluated by the measurement of the conductivity and resonance frequency as a function of time. It is found that during annealing at 700~°C for 350 hours, the resonance frequency of LiNbO$_{3}$ remains in a {\textpm}~100 ppm range of the initial value of 3.5 MHz.

\bigskip

    \section{Introduction}

\bigskip

Lithium niobate (LiNbO$_{3}$, LN) and lithium tantalate (LiTaO$_{3}$, LT) single crystals are among the most studied materials due to their outstanding ferroelectric, piezoelectric, photorefractive, and electro-optical properties and are extensively used for various industrial applications, such as optical waveguides, piezoelectric sensors, optical modulators and many others.\textsuperscript{[1,2]} Both compounds are isostructural (space group \textit{R3c}) with only slight differences in the lattice and positional parameters.\textsuperscript{[3,4]}\ Though the homogeneity range of both LN and LT lies between about 48 mol\% and 50 mol\%Li$_{2}$O
at room temperature, most commercially available crystals are usually grown from a congruently melting composition with an approximate content of about 48.5 mol\% Li$_{2}$O.\textsuperscript{[1,2,5,6]}According to the widely-accepted defect model in congruent
LiNbO$_{3}$ (cLN) the Li-deficiency in the material is compensated via lithium vacancies and antisite defects 
$\text{Nb}_{\text{Li}}^{\text{4{\textbullet}}}${\ .\textsuperscript{[5,7--9]}} The same applies to the congruent LiTaO$_{3}$ (cLT), where the charge neutrality is maintained by formation of antisite $\text{Ta}_{\text{Li}}^{\text{4{\textbullet}}}$ .\textsuperscript{[10,11]} This defect structure controls key physical properties of LN (LT) such as Curie temperature, melting temperature, band gap etc., which strongly depend on Li/Nb (Li/Ta) ratio in the crystal.\textsuperscript{[6,12,13]}

With the composition approaching stoichiometric, the intrinsic defect density is greatly reduced.\textsuperscript{[5,6,12,13]} While cLN and cLT crystals are grown from a melt of congruent composition by the conventional Czochralski method, the growth of homogenous stoichiometric crystals is difficult using this technique due to the incongruent melting of the stoichiometric composition. Large, homogeneous near stoichiometric LiNbO$_{3}$ (nsLN) and LiTaO$_{3}$ (nsLT) crystals can be grown by the high-temperature top-seeded solution growth (HTTSSG) method from K$_{2}$O containing flux \textsuperscript{[13,14]} or from a Li rich melt (58--60 mol\%
Li$_{2}$O) using a modified double crucible Czochralski method with a continuous feeding mechanism.\textsuperscript{[15,16]}
Vapor transport equilibration (VTE) technique is another method to obtain near stoichiometric LN and LT. Here, the off-stoichiometric sample is annealed in a Li-rich atmosphere. In given sufficient time at sufficiently high temperature, the Li/Nb (Li/Ta) ratio in the crystal equilibrates to a desired composition in the powder through a mechanism involving vapor transport and solid-state
diffusion.\textsuperscript{[6,12,13,17]} This technique is limited to thin wafers only, for thicker samples extremely long diffusion times have to be applied.\textsuperscript{[18]}

The application of both LN and LT at elevated temperatures is however limited. LT exhibits a ferroelectric-paraelectric phase transformation at about 630{~°C}.\textsuperscript{[19]}The Curie temperature of LN is about 1200~°C however its high-temperature stability is poor: it has been reported by different authors that LN starts to lose Li$_{2}$O to the environment at temperatures above 300~°C.\textsuperscript{[20,21]}

Recently, attention has been attracted by LiNb$_{1-x}$Ta$_{x}$O$_{3}$ (LNT) solid solutions, which{ can be formed across the whole compositional range and allows materials to be produced with properties that potentially combine}the advantages of end components of the LN{--}LT system.\textsuperscript{[3,4,22,23]} Though the growing process of LNT is extremely challenging due to the differences in the melting temperatures $T_m$ of both end components of the system($T_m$(LiNbO$_{3}$) = 1240~°C; $T_m$(LiTaO$_{3}$) = 1650~°C) and, therefore, the distribution coefficient in the solid/liquid of Ta, $K_{Ta}$ \textgreater \textgreater 1, the growth of sufficiently large LNT crystals and their characterization has been reported by different scientific groups \textsuperscript{[3,4,22,24,25]} as well as by the groups, involved in this study.\textsuperscript{[26--28]}

The current work focuses on the influence of the lithium stoichiometry on the high-temperature electrical and acoustic properties of Li(Nb,Ta)O$_{3}$crystals. To evaluate its impact, as-grown (Li-deficient) and VTE treated (near stoichiometric) specimens are studied.
The Li$_{2}$O-content is determined by the optical method, evaluating the position of the absorption edge. Finally, the conductivity and the acoustic loss of LNT crystals are measured at elevated temperatures during relatively long time periods, aiming to assess the stability of the studied materials.

    \section{Materials and methods}

\bigskip

        \subsection{Crystals and Specimens}

\bigskip

Nominally undoped Czochralski-grown LiNbO$_{3}$, LiNb$_{0.55}$Ta$_{0.45}$O$_{3}$, LiNb$_{0.30}$Ta$_{0.70}$O$_{3}$
and LiTaO$_{3}$ were investigated in this work. LiNbO$_{3}$ and LiTaO$_{3}$ X-cut wafers were supplied by Precision Micro-Optics Inc. (USA).

Single crystalline LNT solid solutions with nominal composition of LiNb$_{0.55}$Ta$_{0.45}$O$_{3}$ and LiNb$_{0.30}$Ta$_{0.70}$O$_{3}$ were grown at the Leibniz-Institut für Kristallzüchtung, Berlin, Germany (IKZ) from a melt consisting of the congruently melting compositions of LiNbO$_{3}$ and LiTaO$_{3}$. High-purity raw materials (4N and better) of Li$_{2}$CO$_{3}$, Nb$_{2}$O$_{5}$, and Ta$_{2}$O$_{5}$ were mixed in the appropriate ratio, pressed at 2 kbars, and sintered at 1000~°C before melting in an iridium crucible. Growth processes were started on oriented seeds and the crystals were pulled with a rate of 0.2 mm/h to suppress cellular growth. The crystals were grown in Ar + 0.08 vol\% O$_{2}$ atmosphere.

Chemical compositions of the crystals were determined on longitudinal sections by X-ray fluorescence (XRF) analysis using a Bruker M4 Tornado spectrometer. Neglecting the Li-stoichiometry, the crystal composition LiNb$_{1-x}$Ta$_{x}$O$_{3}$ was defined as X$_{LT}$ according to:

\begin{equation}
    X_{LT}=\frac{[Ta]}{[Ta]+[Nb]}{\cdot}100\%
\end{equation}

The resulting effective distribution coefficient of about 2.4 is in excellent agreement with the value reported by other authors.\textsuperscript{[22,25]}

Subsequently, the grown LNT crystals were cut into 0.5--0.8 mm thick plates and polished to optical quality. The samples for investigations were then prepared from the plate, cut from the middle part of LNT crystals. Part of the samples was then used for further VTE-treatment, described in details in \textbf{Section 2.2}. The list of the investigated samples, their orientation, Nb/Ta ratio and dimensions are summarized in \textbf{Table 1} along with the type of studies that were employed. In general, the as-grown specimens got the ``c'' prefix, corresponding to the Li-deficient composition, i.e. the congruently melting composition of the end-members LN and LT, while the VTE-treated samples are labeled with ``ns'', referring to their assumed near stoichiometric Li-content.

\bigskip

    {\textbf{Table 1.} Sample labels, compositions, orientations, dimensions and study methods of the investigated LiNb$_{1-x}$Ta$_{x}$O$_{3}$ specimens.}

\begin{center}
\tablefirsthead{}
\tablehead{}
\tabletail{}
\tablelasttail{}
\begin{supertabular}{m{1.6cm}m{0.945cm}m{1.798cm}m{1.5519999cm}m{1.55cm}m{6.5cm}}
\hline
Sample name & X$_{LT}$ [\%] & Orientation & Size [mm$^{2}$] & Thickness [mm] & Study\\
\hline
cLN & 0 & X & 5x6 & 0.5 & absorption edge, conductivity\\
cLN-A & 0 & X & 10 ${\oslash}$ & 0.5 & acoustic loss, long-term\\
nsLN & 0 & X & 5x6 & 0.5 & VTE, absorption edge, conductivity\\
nsLN-A & 0 & X & 10 ${\oslash}$ & 0.5 & VTE, acoustic loss, long-term\\
cLNT45 & 45 & X & 4x6 & 0.8 & absorption edge, conductivity, long-term\\
nsLNT45 & 45 & X & 4x6 & 0.7 & VTE, absorption edge, conductivity\\
cLNT70 & 70 & Y & 5x5 & 0.8 & absorption edge, conductivity, long-term\\
nsLNT70 & 70 & Y & 4x4 & 0.8 & VTE, absorption edge, conductivity\\
cLT & 100 & X & 5x6 & 0.5 & absorption edge, conductivity\\
nsLT & 100 & X & 5x6 & 0.5 & VTE, absorption edge, conductivity\\\hline
\end{supertabular}
\end{center}
${\oslash}$ For acoustic loss measurements 10 mm diameter round samples were used, coated with 5 mm diameter keyhole-shaped Pt electrodes. Rectangular samples were fully covered with Pt electrodes.

\bigskip

        \subsection{Preparation of Nearly Stoichiometric Specimens}

\bigskip

The vapor transport equilibration (VTE) technique was applied to obtain near stoichiometric Li(Nb,Ta)O$_{3}$ samples,  as it was previously reported for lithium niobate \textsuperscript{[12,17]} and tantalate.\textsuperscript{[6,29]} This method involves placing the crystal samples, along with a much larger mass of Li-rich powder with a desired composition and corresponding Nb/Ta ratio and its subsequent annealing at sufficiently high temperatures (\textgreater1000~°C). During this process, the Li/Nb (or Li/(Nb+Ta)) ratio within the specimen equilibrates to a steady-state concentration due to vapor transport mechanism and diffusion.

Two-phase powder mixtures were separately prepared using Li$_{2}$CO$_{3}$, Nb$_{2}$O$_{5}$ and Ta$_{2}$O$_{5}$ combined as starting chemicals, with a Nb/Ta ratio respective to the as-grown Li(Nb,Ta)O$_{3}$ crystals. The powder ratios were selected to yield a net composition of 65 mol\% Li$_{2}$O in a lithium-rich two-phase mixture, composed of Li$_{3}$Nb$_{1-x}$Ta$_{x}$O$_{4}$ and LiNb$_{1-x}$Ta$_{x}$O$_{3}$,which served as the Li$_{2}$O source. Subsequently, these mixtures were heated to 950~°C with a heating rate of 2 K/min and maintained this temperature for 24 hours.

Approximately 15 grams of the resulting lithium-rich mixture were placed in a platinum crucible, along with a sample weighing about 0.2 grams, positioned close to the mixture and supported by alumina rods. The crucible was then loosely closed with a platinum lid. Subsequently, the crucible was heated up to 1100~°C at a rate of 2 K/min and held at that temperature for 168 hours. Note, that this temperature was insufficient to obtain near stoichiometric LiTaO$_{3}$, therefore the VTE procedure was conducted at 1200~°C for this specimen.

\bigskip

    \section{Measurement and Analysis Techniques}

\bigskip

        \subsection{Determination of the Li$_{2}$O Composition}

\bigskip

The Li$_{2}$O content in the samples was determined by the position of the absorption edge. The spectra in the vicinity of the absorption edge were measured with an optical spectrophotometer (Perkin Elmer Lambda 900). The absorption data was corrected for reflection losses applying common equations.\textsuperscript{[30]} The refractive indices were taken from \textsuperscript{[31]} for LN and from \textsuperscript{[32]} for LT. For mixed LiNb$_{1-x}$Ta$_{x}$O$_{3}$ crystals the refractive indices have been calculated as $n_{LNT}=(1-x) n_{LN }+x n_{LT}$.

The wavelength $\lambda_{20}$ read at absorption coefficient $\lambda$ = 20 $cm^{-1}$ has been used to calculate the Li$_{2}$O content according to the empirical equation for LN, suggested in \textsuperscript{[30]}:

\begin{equation}
    \frac{1240}{\lambda _{20}}=-0.189\sqrt{50-c_{\mathit{Li}}}+4.112
\end{equation}

where $c_{Li}$ is the {the Li}$_{2}$O content of the sample in mol\%.

The suggested equation \textsuperscript{[29]} for Li$_{2}$O content determination in LiTaO$_{3}$ crystals is:

\begin{equation}
    \lambda _{20}=276.69-14.18[\exp \left(c_{\mathit{Li}}-50\right)1.51]
\end{equation}

The presence of Ta-ions does not allow for exact determination of Li$_{2}$O content in LNT specimens after VTE treatment, however it still could be estimated from the absorption edge shift, as the dependencies of the absorption edge shift on the Li$_{2}$O content for LiNbO$_{3}$ and LiTaO$_{3}$ are known.

\bigskip

        \subsection{Electrical Conductivity}

\bigskip

The temperature dependent electrical conductivity investigations were conducted on the rectangular plates, described in \textbf{Section 2.1}. Platinum electrodes with a thickness of around 3 {\textmu}m, were applied to both sides of the plates via screen printing. Subsequently, the plates underwent annealing at 900~°C for 1 hour.

The investigations, employing impedance spectroscopy (Solartron 1260), spanned a frequency spectrum from 1 Hz to 1 MHz with an excitation voltage of 50 mV. These measurements were carried out within the temperature range of 400~°C to 900~°C in air at atmospheric pressure with heating at a rate of 1 K/min. The impedance values at temperatures below 400~°C were too high for reliable data acquisition.

The acquired impedance spectra were represented in the complex plane, in form of a Nyquist plot that correlated the imaginary part (Z'') of impedance against the real part (Z'). Subsequently, an electrical equivalent-circuit model, involving a constant phase element (CPE), connected in parallel with a bulk resistance (\textit{$R_B$}), was fitted to the experimental data. The low-frequency intercepts of the \textit{$R_B$}-CPE semicircles in this complex impedance plane were construed as the bulk resistance, which was then transformed into the bulk conductivity $\sigma$. This conversion was achieved using the relation $\sigma =t(A${\texttimes}$R_B)^{-1}$, where \textit{t} denotes the sample thickness, and \textit{A} represents the electrode area. The uncertainty in the conductivity measurements has not exceeded 8 \%.

\bigskip

        \subsection{Acoustic Loss}

\bigskip

Acoustic loss investigations were carried out by resonant piezoelectric spectroscopy (RPS) on X-cut resonators operating in the thickness-shear mode (TSM). The samples were mounted in an aluminum oxide sample holder and electrically contacted using platinum foils on each side of the sample. To minimize mechanical damping from the contact, only small areas near the sample edges were electrically contacted and mechanically clamped, employing keyhole-shaped platinum electrodes, deposited as described in \textbf{Section 3.2}. These electrodes were designed to overlap in a circular pattern at the non-clamped center of the samples. The measurements were performed in tube furnace in air at atmospheric pressure while heating at a rate of 1 K/min from ambient temperature.

The real and imaginary parts of the impedance spectra close to the resonance frequency were recorded using the network analyzer (HP E5100A). Subsequently, the data were converted into admittance $Y=Z^{-1}$.
By fitting a Lorentz function to the resonant peak of the real part of $Y$, the resonant frequency was determined. From this data, the inverse quality factor, $Q^{-1}$ was calculated, as a measure of acoustic loss. Detailed description of data acquisition is given in \textsuperscript{[33]}.

\bigskip

    \section{Results and Discussion}

        \subsection{Li-content}

\bigskip

The absorption spectra of congruent and VTE-treated specimens in the vicinity of the absorption edge are shown in \textbf{Figure 1}, plotted separately for the end members of the LNT system (Figure 1.a) and for the LNT solid solutions (Figure 1.b). As seen
from Figure 1.a), the absorption coefficient equals 20 cm$^{-1}$ for cLN and cLT at 319.5 nm and at 275.5 nm, respectively. These values are in a good agreement with those, reported by other research groups for congruent LiNbO$_{3}$ \textsuperscript{[30,31]} and LiTaO$_{3}$.\textsuperscript{[30]} The absorption edge of the VTE-treated specimens are shifted towards the deep UV range, compared to the as-grown crystals (Figure 1). The wavelengths, at which the absorption coefficient reaches 20~cm$^{-1}$, used for the determination of Li$_{2}$O content are summarized in \textbf{Table 2}.

    {\ } \includegraphics[width=7.281cm,height=4.944cm]{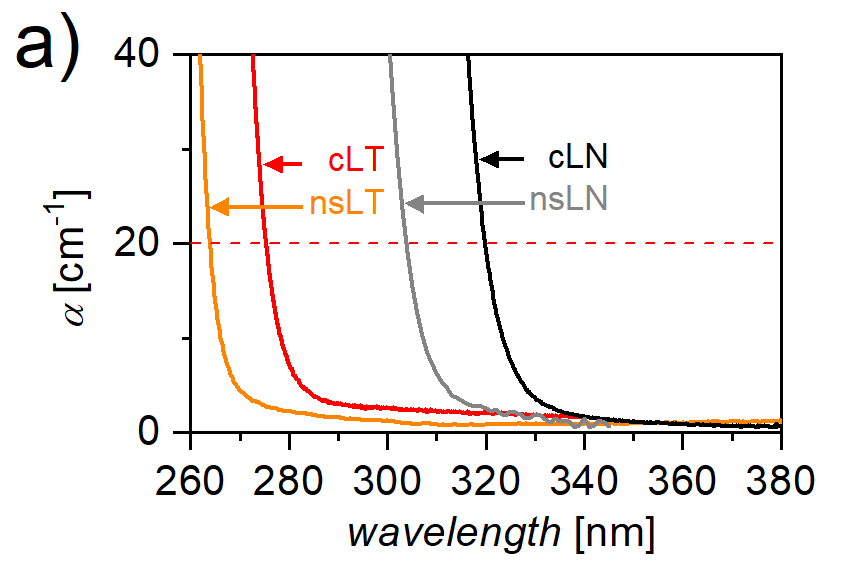}
    {\ } \includegraphics[width=7.281cm,height=4.944cm]{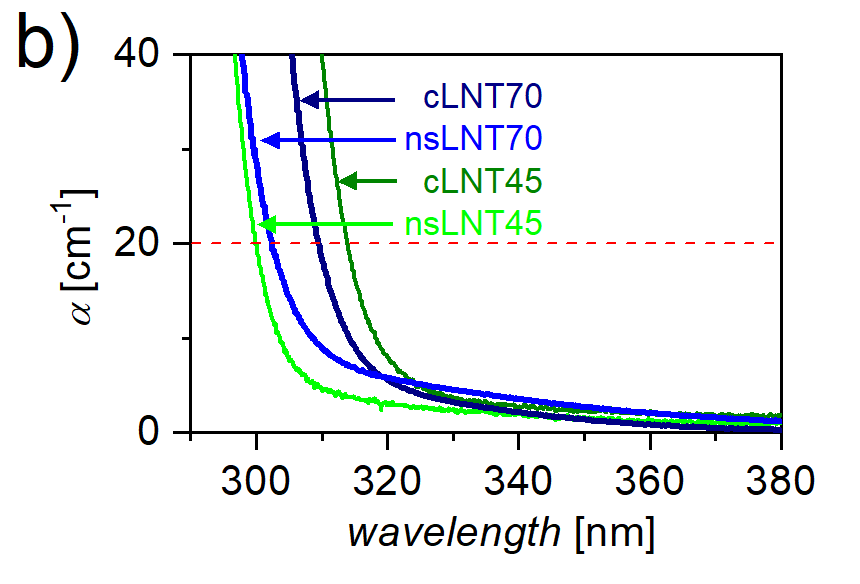}
    
    {\textbf{Figure 1} Absorption spectra near the UV edge of the as-grown and near stoichiometric specimens of a) LN and LT and b) LNT45 and LNT70.}

\bigskip

    \textbf{Table 2}. Wavelengths of the absorption edge read at $\lambda$ = 20 $cm^{-1}$ of the as-grown and VTE-treated LNT samples.

\begin{center}
\tablefirsthead{}
\tablehead{}
\tabletail{}
\tablelasttail{}
\begin{supertabular}{m{3.215cm}m{3.12cm}m{3.1599998cm}}
\hline
\multicolumn{1}{m{3.215cm}|}{~} &
\multicolumn{2}{m{6.48cm}}{\centering{{Wavelength @ ${\alpha}$=20 cm$^{-1}$ [nm]}}}\\\hhline{~--}
\multicolumn{1}{m{3.215cm}|}{Sample} & \multicolumn{1}{m{3.12cm}|}{as-grown} & {{VTE treated}}\\\hline
LN & 319.5 & 303.5\\
LNT45 & 313.5 & 300.0\\
LNT70 & 309.5 & 303.0\\
LT & 275.5 & 263.5\\\hline
\end{supertabular}
\end{center}

\bigskip

The calculated Li$_{2}$O content in VTE-treated samples (\textbf{Equation 2} and \textbf{3}) yields more than 49.9 mol\% for both nsLN and nsLT specimens, which is already very close to the nominal stoichiometry.

Further, as seen from Figure 1, the introduction of Ta shifts the absorption edge of the crystal towards smaller wavelengths, which does not allow for a precise determination of Li$_{2}$O content in LiNb$_{0.55}$Ta$_{0.45}$O$_{3}$ and LiNb$ _{0.30}$Ta$_{0.70}$O$_{3}$.
The latter, however, could be estimated, using the dependencies of the absorption edge shift on the Li$_{2}$O content for LiNbO$_{3}$ and LiTaO$_{3}$. As shown in \textsuperscript{[29,30]}, the change of Li$_{2}$O stoichiometry from 48.4 mol\% (congruent composition to 50 mol\% (stoichiometric composition) leads to the shift of absorption edge towards smaller wavelength of about 16 nm for LiNbO$_{3}$ \textsuperscript{[30]} and of about 13 nm for LiTaO$_{3}$.\textsuperscript{[29]}
Therefore, using the expressions (2) and (3) it can be calculated, that the UV absorption edge shift of 6.5 nm,
observed for the nsLNT70 specimen, corresponds to approximately 49.4 mol\% Li$_{2}$O in LN (Equation 2) and to 49.6 mol\% in LT
(Equation 3). Their weighted average for LiNb$ _{0.30}$Ta$_{0.70}$O$_{3}$, resulted in an estimated Li$_{2}$O content exceeding 49.5 mol\%.

Similarly, for LNT45, a 13.5 nm shift in both LN and LT implied a Li$_{2}$O content over 49.9 mol\%. Consequently nsLNT45 is presumed to contain at least 49.9 mol\% Li$_{2}$O.

\bigskip

        \subsection{Electrical Conductivity}

\bigskip

{The temperature dependent electrical conductivity of near stoichiometric and Li-deficient Li(Nb,Ta)O$_{3}$ samples is presented in
\textbf{Figure 2} in the form of an Arrhenius plot. The measurements were performed in air in the temperature range from about 400--450~°C to 900~°C for all the studied samples, except for the specimen nsLNT45, which was studied above 650~°C. As seen from Figure 2 generally the near stoichiometric samples exhibit lower conductivity, compared with Li-deficient Li(Nb,Ta)O$_{3}$. This result is expected and attributed to the differences in the defect subsystem of the Li-deficient samples and those, subjected to the VTE
treatment. According to the studies of the electrical properties in single crystalline LN,\textsuperscript{[34--39]} LT \textsuperscript{[40--42]} and recently also in LNT solid solutions,\textsuperscript{[27,28,43,44]} the ionic conduction
mechanism via lithium vacancies governs the conductivity in these materials in the range from at least 300~°C to about 600--700~°C. As the temperature increases, the defects are thermally activated and the mobility of the lithium ions is facilitated. In near stoichiometric LNT samples, however, the number of Li vacancies is greatly reduced, which on the other hand also suppresses the ionic conduction. This assumption is in line with the Li-diffusivity studies in LiNbO$_{3}$, performed in \textsuperscript{[45]}, where the authors show that the diffusivity of Li in near stoichiometric samples is reduced by about one order of magnitude up to 600~°C compared to the congruent material. We note also that the lower electrical conductivity of near stoichiometric LiNbO$_{3}$, comparing to the congruent specimens was observed previously in \textsuperscript{[34]} consistent with the current study.}

    {\centering  \includegraphics[width=9.984cm,height=7.938cm]{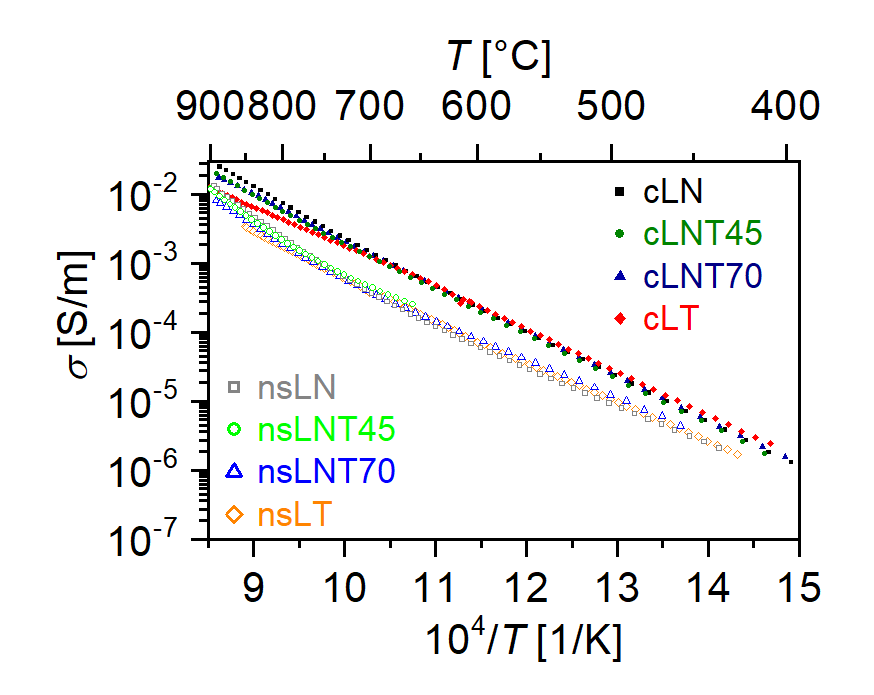} \par}
    \textbf{Figure 2.} Temperature-dependent conductivity of different Li(Nb,Ta)O$_{3}$ samples.

\bigskip

{Given that ionic transport dominates, the conductivity \textit{$\sigma $} has a form:}

\begin{equation}
    \sigma =\frac{\sigma _0} T\exp \left(\frac{-E_A}{kT}\right)
\end{equation}

where, $\sigma_0$ is a pre-exponential constant, $T$ is an absolute temperature, $E_A$ is an activation energy and $k$ is the Boltzmann constant. Considering that the Li-ionic conduction mechanism is a diffusive transport process, the conductivity becomes inversely proportional to the temperature and \textbf{Equation 4} includes term 1/T, reflecting the relationship between charge carrier mobility and thermally activated diffusion process, according to the Nernst-Einstein relation. Accordingly, the product $\sigma T$ must be used to determine the activation energy, provided that a single conduction mechanism dominates in a sufficiently large temperature range.

{Further, as seen from Figure 2, the temperature-dependent conductivities do not fully obey the Arrhenius law in the
investigated temperature range, which must be evaluated in detail. Consequently, the related slopes of \textit{$\sigma
$T} are calculated for the studied here samples using the expression \textsuperscript{[39]}:}

\begin{equation}
    E_{\sigma T}=-k\frac{{\partial}\left(\ln \sigma T\right)}{{\partial}\left(1/T\right)}
\end{equation}

Here $E_{\sigma T}$ corresponds to the activation energy with $E_A = E_{\sigma T}$, provided that a single conduction mechanism dominates over considerably wide temperature range. \textbf{Figure 3} represents the calculated slopes $E_{\sigma T}$, plotted separately for Li-deficient (Figure 3.a) and near stoichiometric (Figure 3.b) Li(Nb,Ta)O$_{3}$ specimens. As seen from Figure 3.a), the slope is found nearly constant for all the studied Li-deficient samples up to about 600--650~°C, depending on samples composition. This enables the evaluation of activation energies for this temperature range by averaging of the $E_{\sigma T}$: $E_A$ = (1.30 {\textpm} 0.07), (1.21 {\textpm} 0.06), (1.25 {\textpm} 0.06) and (1.31 {\textpm} 0.07) eV for cLN, cLNT45, cLNT70 and cLT specimens, respectively.

    {\centering  \includegraphics[width=7.886cm,height=6.16cm]{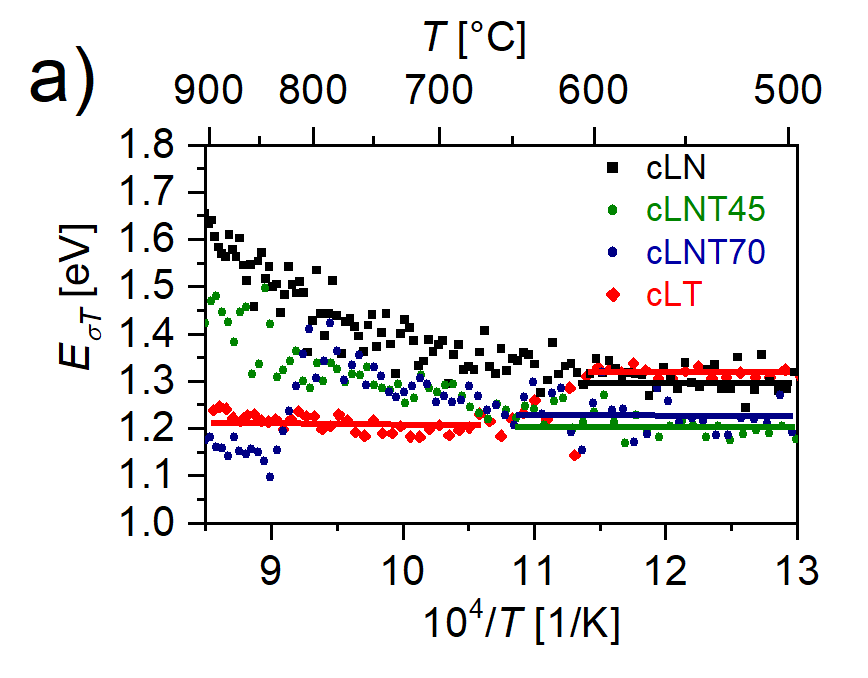} 
    \includegraphics[width=7.932cm,height=6.197cm]{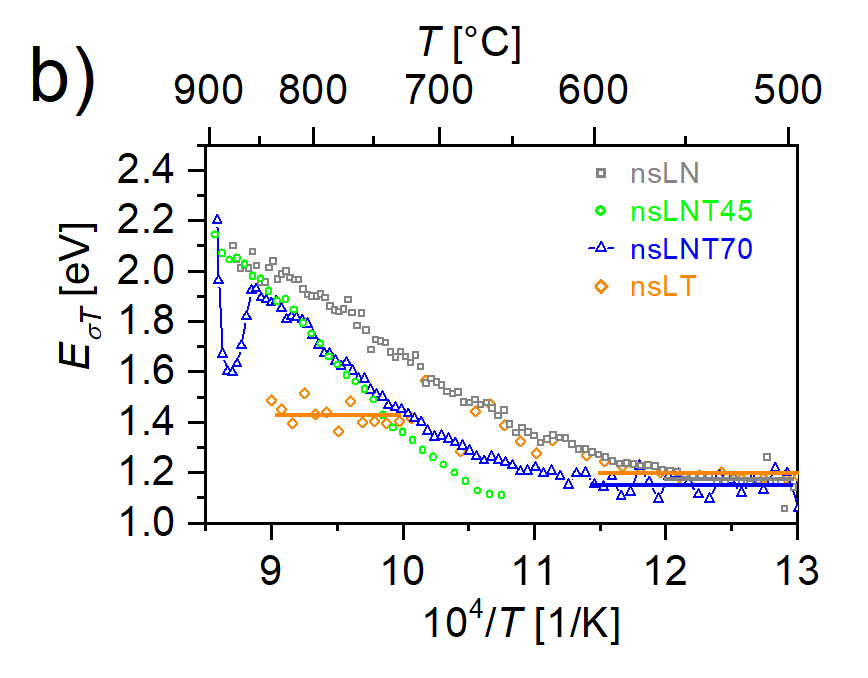} \par}
    \textbf{Figure 3.} Slopes $E_{{\sigma}T}$ of the electrical conductivity of Li-deficient (3.a) and near stoichiometric (3.b) LNT samples (symbols) and related ranges of approximately constant derivatives (lines).

\bigskip

{The constant range of the calculated slopes for the near stoichiometric samples (Figure 3.b) is found up to 550--600~°C. Here the extracted activation energies are equal to $E_A$ = (1.18 {\textpm} 0.05), (1.16 {\textpm} 0.05) and (1.20 {\textpm} 0.06) eV for nsLN, nsLNT70 and nsLT specimens, respectively. No constant range was observed for the nsLNT45, while the measurements for this specimen were performed only above 650~°C, as mentioned earlier. The determined values of activation energies are consistent with those, obtained in other studies for Li-ion conduction mechanism in LN \textsuperscript{[34--39]}, LT \textsuperscript{[40--42]} as well as in our studies of LNT solid solutions.\textsuperscript{[27,28,43,44]}}

Above 550--650~°C the calculated slopes are no longer constant and their behavior depends on the composition of the studied samples. The cLT specimen shows slope decrease at about 620--630~°C (Figure 3.a), which corresponds to the Curie temperature $T_C$ of congruent LT. Above 650~°C the slope becomes nearly constant again, which allows the determination of activation energy for this region:
$E_A$ = (1.22 {\textpm} 0.06) eV. An analogous drop of the activation energy was observed for the cLNT70 at about 800--820~°C, which also corresponds to the ferroelectric-paraelectric phase transformation in this specimen. A similar behavior was reported previously in \textsuperscript{[39,44]} for LT and Ta-rich LNT specimens and attributed to the peculiar distribution of the Li-ions and Li-vacancies on the octahedral sites, which are involved in the phase transformation near $T_C$.

The Curie temperature in near stoichiometric crystals is shifted towards higher temperatures, compared to the Li-deficient Li(Nb,Ta)O$_{3}$, provided that the Nb/Ta ratio is identical.\textsuperscript{[6,18]} Thus, the slope $E_{{\sigma}T}$, calculated for the nsLT specimen shows a gradual increase in the range of about 600--670~°C. Subsequently, a transition range of 670--700~°C is observed, which is related to $T_C$ in nsLT, above which the slope becomes nearly constant again, however, in contrast to cLT, the values of $E_{{\sigma}T}$ are higher in this range: $E_A$ = (1.42 {\textpm} 0.07) eV. For the nsLNT70 specimen the drop of $E_{{\sigma}T}$, related to phase transformation,is registered at about 850~°C, followed by a transition range 850--890~°C, which is already the highest measured temperature for this sample.

It can be observed for LT and LNT70 pairs, that the increase of the Curie temperature from the Li-deficient to the near stoichiometric specimen is over 70~°C, which is comparable to the shift of approximately 90~°C in LN \textsuperscript{[46]} and in LT
\textsuperscript{[47]} measured by other methods.

Obviously, different processes govern the conductivity below and above the $T_C$ in near stoichiometric and Li-deficient LNT specimens. The study of conduction mechanisms in paraelectric phase is however beyond the scope of current work and is discussed in separate
publications.\textsuperscript{[45,48]}

Further, a gradual increase of the slope with increasing temperature is observed for cLN, cLNT45 and cLNT70 above 600--650~°C (Figure 3.a). For the near stoichiometric samples the slope increase is even more pronounced and reaches the value of 2.1--2.2 eV at 900~°C (Figure 3.b).
This implies that a different process increasingly contributes to the conductivity with increasing temperature, which does not
allow assigning a single activation energy for this range. Note, that for cLNT70 and for nsLNT70 the $E_{{\sigma}T}$ increase overlaps with $T_C$,while for cLN and cLNT45 as well as for nsLN and nsLNT45 the $T_C$ is not reached.

{In order to separate the high-temperature contribution, the low-temperature conductivity, derived by fitting Equation 4 to the measured data up to 550--650~°C,is extrapolated to the upper limit of the measured range and then subtracted from the dataset. The range for the fit was chosen individually for each sample to ensure that the corresponding slope remains constant (Figure 3). This approach was employed in a previous work \textsuperscript{[44]} for Li-deficient LNT, where the high-temperature contribution was attributed to the electronic conduction mechanism, which will be discussed below. In current study, this approach is exemplarily illustrated in \textbf{Figure 4} for the nsLN specimen. Here the solid gray line represents a fit of Equation 4 to the data in the range from 430~°C to 560~°C { and its extrapolation to 900}~°C. The blue squares in Figure 4 show the high-temperature contribution, which is the difference between the measured data and the extrapolated low-temperature contribution, prompted by Li-ion migration. Subsequently, Equation 4 is fitted to the evaluated difference and the fit result is also plotted in Figure 4 as a straight line in light blue color. The
activation energy obtained from this fit yields (2.35 {\textpm} 0.09) eV.}

{Further, given that the ionic conduction mechanism is dominant at low temperatures whereas at high temperatures two different mechanisms simultaneously contribute the overall electrical conductivity, the latter can be written as:}

\begin{equation}
    \sigma =\frac{\sigma _{LT}} T\exp \left(\frac{-E_{ALT}}{kT}\right)+\frac{\sigma _{HT}} T\exp \left(\frac{-E_{AHT}}{kT}\right){.}
\end{equation}

Here, the first and the second terms of the \textbf{Equation 6} represent the low-temperature and high-temperature contribution to
\textit{{$\sigma$}}, respectively, where $\sigma_{LT}$ and $\sigma_{HT}$ are the pre-exponential constants for the ionic and electronic conduction mechanisms and $E_{ALT}$ and $E_{AHT}$ -- their respective activation energies.

    {\centering  \includegraphics[width=8.601cm,height=7.001cm]{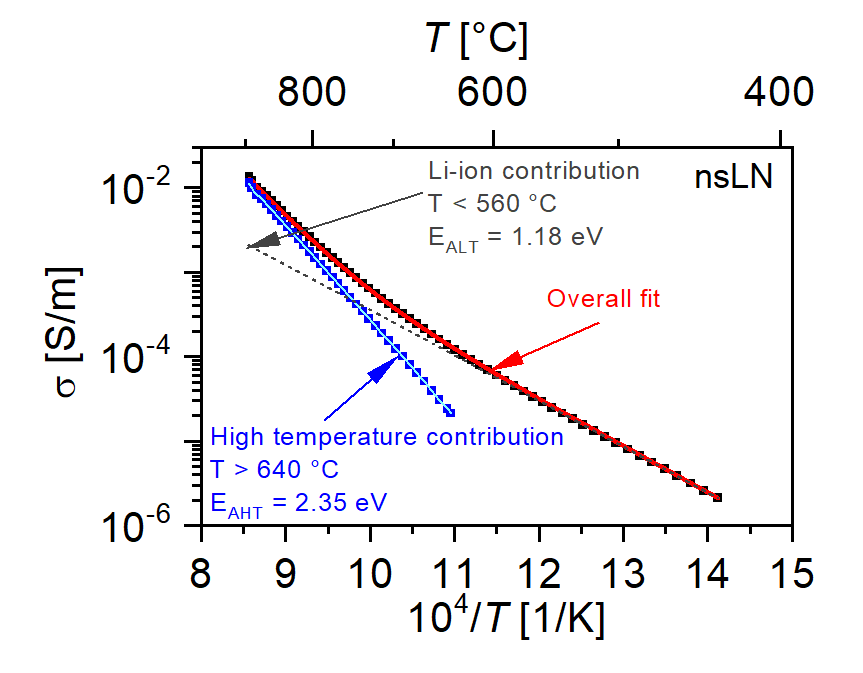} \par}
    \textbf{Figure 4} Temperature-dependent conductivity of nsLN specimen with fit according to Equation {6} and separated low- and high-temperature contributions to conductivity (see text for details).

\bigskip

Equation 6 was employed to fit the measured conductivity of the investigated samples. As discussed above, the activation energy of Ta-rich samples is affected by the Curie temperature, thereby imposing constraints on data interpretation. To mitigate this influence, the fitting procedure for the samples was conducted within temperature ranges where there are no apparent disruptions induced by the structural phase transformation at the $T_C$. Accordingly, fits for cLT and nsLT specimens were performed above $T_C$, considering the negligible impact of HT-contribution on electrical conductivity below this threshold. For all other samples the fit was performed below $T_C$. The resulting activation energies, derived from the fitting Equation 6 to the measured data are listed in \textbf{Table 3} along with the temperature ranges, chosen for evaluation. The fit result is exemplarily shown for nsLN sample in Figure 4. As seen from Figure 4 the fit outcome (red line) exhibits excellent agreement with the measured dataset (represented by black squares).

\bigskip

    \textbf{Table 3}. Activation energies of conductivity calculated according to Equation 6.

\begin{center}
\tablefirsthead{}
\tablehead{}
\tabletail{}
\tablelasttail{}
\begin{supertabular}{m{2.271cm}m{1.5799999cm}m{0.8cm}m{0.29900002cm}m{1.8cm}m{2.8cm}m{2.792cm}}
\hline
\multicolumn{1}{m{2.271cm}|}{~} &
\multicolumn{4}{m{5.0790005cm}|}{\centering{Temperature range [°C]}}
&
\multicolumn{2}{m{5.7920003cm}}{\centering{Activation energy [eV]}}\\\hhline{~------}
\multicolumn{1}{m{2.271cm}|}{Sample } & ~ & ~ &
\multicolumn{2}{m{2.299cm}|}{~} &
\multicolumn{1}{m{2.8cm}|}{$E_{ALT}$} & {$E_{AHT}$} \\\hline
nsLN & {{430}} & \multicolumn{2}{m{1.299cm}}{$-$} & 890 & {1.18 {\textpm} 0.05} & {2.35 {\textpm} 0.09}\\
nsLNT45 & {650} & \multicolumn{2}{m{1.299cm}}{$-$} & 900 & {1.08 {\textpm} 0.05} & 2.46 {\textpm} 0.20\\
nsLNT70 & {460} & \multicolumn{2}{m{1.299cm}}{$-$} & 850 & {1.15 {\textpm} 0.05} & 2.30 {\textpm} 0.45\\
nsLT & {710} & \multicolumn{2}{m{1.299cm}}{$-$} & 850 & {1.40 {{\textpm} 0.06}} & 2.10 {\textpm} 0.90\\
cLN & 430 & \multicolumn{2}{m{1.299cm}}{$-$} & 900 & {1.30 {\textpm} 0.07} & 2.20 {\textpm} 0.08\\
cLNT45 & 440 & \multicolumn{2}{m{1.299cm}}{$-$} & 890 & {1.22 {\textpm} 0.05} & 2.32 {{\textpm} }{0.34}\\
cLNT70 & 400 & \multicolumn{2}{m{1.299cm}}{$-$} & 770 & {1.25 {\textpm} 0.05} & 2.39 {{\textpm} 0.70}\\
cLT & 670 & \multicolumn{2}{m{1.299cm}}{$-$} & 900 & {1.20 {{\textpm} 0.05}} & 2.10 {{\textpm} 0.90}\\\hline
\end{supertabular}
\end{center}

\bigskip

The activation energies for low-temperature range, derived from fit of Equation 6 (Table 3), closely align with the values obtained through the averaging of $E_{{\sigma}T}$ (Figure 3), as expected. For high-temperature contribution, the activation energies ranging from 2.1 eV to 2.5 eV were determined, and their discussion in the context of potential conduction mechanisms will follow. \ Note that the uncertainty increases with increasing Ta content as the difference between the bulk conductivity and low-temperature contribution becomes smaller.

Analysis of literature suggests that at sufficiently elevated temperatures (e.g. above 700--800~°C) niobium and/or tantalum ions may potentially diffuse in Li(Nb,Ta)O$_{3}$ through the Li-vacancies, given the similar ionic radii of Nb, Ta and Li (r(Nb$^{5+}$) = r(Ta$^{5+}$) = 64 pm; r(Li$^{+}$)~=~76~pm).\textsuperscript{[49]}
The Nb- and Ta- diffusivities in LiNbO$_{3}$ were studied in \textsuperscript{[50]}, where the activation energy of 2.75 eV was determined for the Nb-diffusion in congruent LN in the temperature range of 1000--1100~°C. This value aligns relatively closely with the activation energies obtained for high-temperature contribution in the present study.
However our earlier research \textsuperscript{[28]} estimated the Nb-ionic conductivity in LN based on Nb-diffusion coefficients, published in \textsuperscript{[50]}, revealing that at 900~°C it does not exceed 0.01 \% of total bulk conductivity in congruent LN, which does not allow attributing the high-temperature contribution solely to Nb migration. The same applies to the transport of oxygen, which can be disregarded due to the much higher activation energy of O-diffusion of about 3.4 eV \textsuperscript{[17,51]} and significantly lower diffusion coefficients compared to those of lithium.\textsuperscript{[17,52]}

Moreover, there is potential for hydrogen to contribute to electrical conductivity, given its high mobility and possible presence in LNT in the form of OH-groups. It, however, is expected to leave the lattice at elevated temperatures, so its contribution decreases as the temperature increases. In addition, the activation energies of about 1 eV are found for H-diffusion process in LN at temperatures well below the range, considered in present investigation, as reported in \textsuperscript{[53,54]}. This finding is not consistent with the values for high-temperature contribution, obtained here.

Further, as already mentioned above, in the study by Yakhnevych et al. \textsuperscript{[44]} it was shown that the rise of the slope of the electrical conductivity with temperature in Li-deficient Li(Nb,Ta)O$_{3}$ is attributed to the increase of the electronic conduction. This finding is in line with the investigations, presented in \textsuperscript{[34]}, which reports an ionic transference number of only 70\% for LiNbO$_{3}$ at 1000~°C in ambient atmosphere. Therefore, the electronic contribution could reach up to 30\%, substantiating the aforementioned statement. Further, the electronic conductivity with activation of 2.5 eV at high temperatures is reported in \textsuperscript{[55]} for LN, where the transport is assigned to the hopping mechanism of small polarons. Additionally, the activation energies between 2.1 eV and 2.2 eV, reported in \textsuperscript{[56,57]} for LN were attributed to the electronic conduction.

Moreover, at elevated temperatures the concentration of lithium vacancies in LNT increases with temperature, due to the Li$_{2}$O out-diffusion and evaporation, which may lead to the increase of electronic conductivity with polarons being involved.\textsuperscript{[58]} In VTE-treated LNT this effect is even more pronounced, as here the Li-ion conduction mechanism is suppressed due to the reduced number of Li-vacancies and the contribution of the electronic conduction is more clearly visible compared to Li-deficient LNT. This finding is
indirectly confirmed by the increase of the slope $E_{{\sigma}T}$, which for the near stoichiometric samples is much stronger, compared to the Li-deficient LNT (Figure 3).

\bigskip

        \subsection{Acoustic Loss}

\bigskip

The inverse quality factor $Q^{-1}$ (acoustic loss) of congruent LiNbO$_{3}$ and VTE-treated LiNbO$_{3}$ (samples cLN-A and nsLN-A) is shown in \textbf{Figure 5} as a function of inverse temperature, measured from approximately 200~°C to about 900~°C. The frequencies of the measurements, indicated in the figure, correspond approximately to the fundamental mode of the resonators, operated in the thickness-shear mode. As seen from Figure 5, the overall losses, determined for nsLN-A specimen are somewhat lower, comparing to the congruent cLN-A. We note, however, that the $Q^{-1}$ in LNT exhibits non-identical absolute values in different samples below 600-650~°C, as shown in our previous studies.\textsuperscript{[27,44]} The non-material losses, arising from mounting and electrodes could contribute to the overall $Q^{-1}$ significantly in this temperature range, potentially affecting the $Q^{-1}$ magnitude.

    {\centering  \includegraphics[width=9.061cm,height=6.542cm]{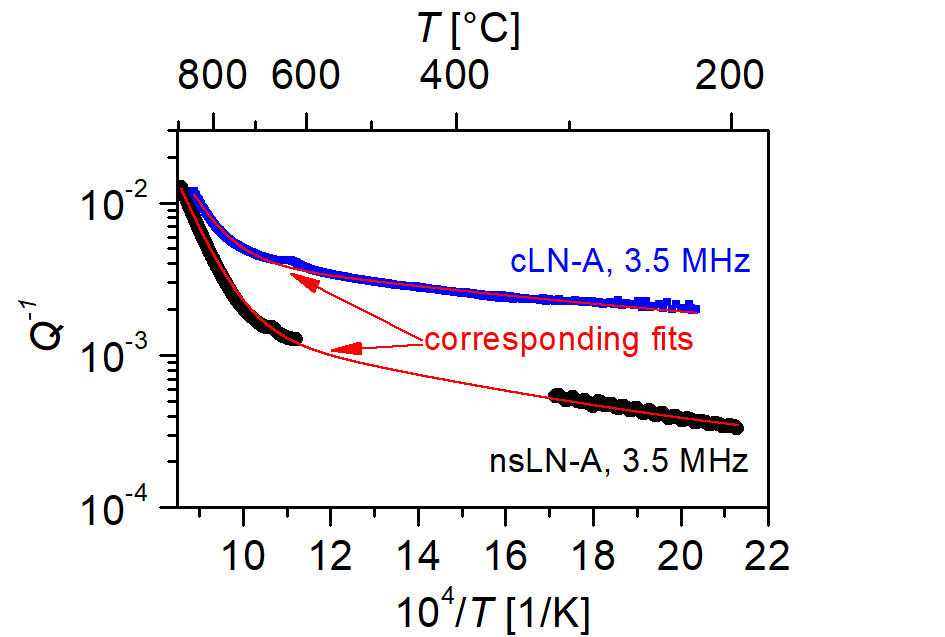} \par}
    \textbf{Figure 5.} Loss in congruent and VTE-treated LiNbO$_{3}$ as a function of inverse temperature and corresponding fits (see text for details).

\bigskip

Further, in the raw data of the nsLN-A sample a significant presence of sharp peaks was detected within the temperature range of approximately 300~°C to 600~°C. These peaks are understood to originate from the phenomenon known as {\textquotedbl}activity dips{\textquotedbl}, arising from the coupling of the primary mode with other unintended (parasitic) modes.\textsuperscript{[59]} This
coupling introduces inaccuracies in the correct determination of the Q-factor. Consequently, the $Q^{-1}$ data corresponding to the temperature range exhibiting activity dips were excluded from the results, plotted in Figure 5 to
ensure a more reliable and clear presentation of the measured data.

Further, the measurements reveal that the $Q^{-1}$ of both LiNbO$_{3}$ samples is monotonically increasing with the temperature up to about 600--650~°C. Such monotonic increase of $Q^{-1}$ has also been reported for a number of piezoelectric crystals \textsuperscript{[60--64]} and will be discussed later in this section. Thus, the measured values of $Q^{-1}$ for cLN-A at 200~°C and 650~°C are equal to about 2.0${\times}$10$^{-3}$ and 4.2${\times}$10$^{-3}$, respectively. As already stated above, the measured $Q^{-1}$ data of nsLN-A sample could not be evaluated in the range of 300--600~°C, however the monotonic increase is also assumed for this nsLN-A specimen (Figure 5). Values of $Q^{-1}$ for nsLN-A at 200 °C and at 600~°C were determined to be 3.45${\times}$10$^{-4}$ and 1.4${\times}$10$^{-3}$, respectively.

Above 650~°C the $Q^{-1}$ of LiNbO$_{3}$ specimens is found to increase even more strongly reaching the value of 1.19${\times}$10$^{-2}$ at 853~°C for cLN-A and 1.27${\times}$10$^{-2}$ at 892~°C for nsLN-A. This rapid loss increase is understood to originate from piezoelectric/conductivity relaxation mechanism, which is expected to contribute substantially to $Q^{-1}$ at megahertz frequencies in piezoelectric materials.\textsuperscript{[61,65]} This contribution is approximately given by:\textsuperscript{[65]}

\begin{equation}
    Q_c^{-1}(\omega ,T){\approx}K^2\frac{\omega \tau _c}{1+\omega ^2\tau_c^2}
\end{equation}

where $\omega$ is the angular frequency (equal to $2$$\pi$$f$); $T$ and $K^{2}$ are the absolute temperature and the electromechanical coupling coefficient, respectively and $\tau$$_c$ is the relaxation time, which is inversely proportional to conductivity with $\tau$$_c$ = $\varepsilon_{ij}/\sigma$.
Here, $\sigma$ and $\varepsilon_{ij}$ denote the electrical conductivity and the dielectric permittivity, respectively. For the thickness-shear mode of Y-cut crystals from point group $3m$,the electromechanical coupling coefficient is equal to $e_{15}^2/(C_{44}\varepsilon _{11})$ in reduced-index notation, where $e_{15}$, $C_{44}$ and $\varepsilon$$_{11}$ are the piezoelectric coefficient, the elastic stiffness and the dielectric permittivity, respectively.\textsuperscript{[66]}

Thereby, the following expression can be written to describe the overall losses $Q^{-1}$ of LiNbO$_{3}$ resonators:

\begin{equation}
    Q^{-1}=K^2\frac{\omega \tau _c}{1+\omega ^2\tau _c^2}+\Delta _T\exp \left(\frac{-E_T}{kT}\right)+C_0
\end{equation}

where

\begin{equation}
    \tau _c=\frac{\varepsilon _{11}}{\sigma }=\frac{\varepsilon _{11}T}{\sigma _0}\exp \left(\frac{E_c}{kT}\right)
\end{equation}

The first term of the \textbf{Equation 8} describes the contribution of piezoelectric/carrier relaxation. The second term represents
the temperature-dependent background that increases monotonically with temperature, where $\Delta _T$ and $E_T$ are the pre-exponential constant and activation energy, respectively. The term $C_0$ describes the temperature independent background, which arises from cables, mounting, electrodes etc. Equation 8 is fitted to the measured $Q^{-1}$ values for cLN-A and nsLN-A and the fits are visualized in Figure 5. In this fit approach the value of $K^{2}$ = 0.57 was calculated, using the material constants of LiNbO$_{3}$ at 600~°C, published in \textsuperscript{[67]}: $e_{15}$ = 4.32 C/m$^{2}$; $C_{44}$ = 53.5~GPa; {$\varepsilon _{11}$} = 6.02{\texttimes}10$^{-10}$ F/m. Subsequently, the $K^{2}$ was approximated as a constant with an aim to obtain stable fit results. The fit parameters are summarized in \textbf{Table 4.}

\bigskip

    \textbf{Table 4}. Fit parameters obtained from fits of $Q^{-1}(T)$ of the cLN-A and nsLN-A samples.

\begin{center}
\tablefirsthead{}
\tablehead{}
\tabletail{}
\tablelasttail{}
\begin{supertabular}{m{2.292cm}m{4.557cm}m{4.552cm}}
\hline
\multicolumn{1}{m{2.292cm}|}{~
} &
\multicolumn{1}{m{4.557cm}|}{\centering{{\textbf{cLN-A}}}} & \centering\arraybslash{{\textbf{nsLN-A}}}\\\hline

\boldmath{$f$} & {{3.5 MHz}} & {{3.5 MHz}}\\
\boldmath{$K^{2}$} & {{0.57 (fixed)}} & {{0.57 (fixed)}}\\
\boldmath{$E_c$} & {{(1.45 {\textpm} 0.07) eV}} & {{(1.53 {\textpm} 0.07) eV}}\\
\boldmath{$\sigma_0$} & {{(5.75 {\textpm} 0.07){\texttimes}10}{\textsuperscript{5}}{ SK/m}} & {{(1.16 {\textpm}
0.08){\texttimes}10}{\textsuperscript{6} SK/m}}\\
\boldmath{$\Delta_T$} & (8.90 {\textpm} 0.05){\texttimes}10$^{-3}$ & {{(5.66 {\textpm} 0.05){\texttimes}10$^{-3}$}}\\
\boldmath{$E_T$} & (0.10 {\textpm} 0.04) eV & (0.14 {\textpm} 0.04) eV\\
\boldmath{$C_0$} & {(1.2 {\textpm} 0.1){\texttimes}10$^{-3}$} & (1.8 {\textpm} 0.1){\texttimes}10$^{-3}$\\\hline
\end{supertabular}
\end{center}

\bigskip

The activation energies for piezoelectric and conductivity relaxation in cLN-A and nsLN-A samples, derived from the $Q^{-1}(T)$ fit, are approximately 1.45 eV and 1.53 eV, respectively. These values are higher, comparing to the activation energies, determined for the lithium conduction mechanism from conductivity measurements (Table 4). However, as discussed in \textbf{Section 4.2}, the conductivity in LiNbO$_{3}$ at the highest-measured temperatures is not solely governed by a single thermally activated process. The apparent increase of $E_{\sigma T}$, determined from the conductivity studies for cLN-A and nsLN-A above 600~°C (see Figure 3) may account for the observed deviations in the evaluated activation energies. Conducting simultaneous loss studies at the fundamental mode and higher harmonics could potentially allow for more precise determination of activation energies, which would be the subject of the subsequent investigations.

Further, as stated above the monotonic increase of $Q^{-1}$, which obeys Arrhenius behavior, observed in current work for LiNbO$_{3}$ specimens have also been reported for a number of piezoelectric crystals.\textsuperscript{[60--64]}
Despite the apparent commonness of such high-temperature background loss, its physical origins are not well established and may have fundamentally different natures in different materials. Martin and Lopez \textsuperscript{[60]} observed high-temperature background loss that increases approximately exponentially with temperature in alkali-compensated quartz crystals. They concluded that this effect arises from diffusion of alkali impurities (Na and Li) after thermally-activated escape from interstitial sites, and this hypothesis is supported by their observation that the temperature-dependent backgrounds are greatly reduced when Na or Li impurities are replaced with hydrogen. Kogut et al \textsuperscript{[64]} observed a similar Arrhenius background loss with an activation energy of 0.22 eV in AlN crystals with high concentrations of oxygen impurities. The authors in \textsuperscript{[61]} have suggested that the high-temperature background, observed in LGS (La$_{3}$Ga$_{5}$SiO$_{14}$) and LGT (La$_{3}$Ga$_{5.5}$Ta$_{0.5}$O$_{14}$) piezoelectric resonators can arise from anelasticity of growth-related structural defects.

The nature of physical mechanisms, responsible for the exponentially increasing background loss in LiNbO$_{3}$ is still to be determined. The frequency dependent measurements, performed for higher overtones, would potentially aid in the determination of these
mechanisms, and will be carried out in subsequent studies.

        \subsection{Long-Term Behavior}

The evaluation of Li(Nb,Ta)O$_{3}$ operational limits involved an examination of electrical conductivity and resonance frequency over a period exceeding 2 weeks (350 hours) at high temperatures in an air environment. These studies were performed for cLN-A and nsLN-A specimens at 700~°C. As seen from the Figure 6, the fundamental resonance frequency of both measured samples shows noticeably stable behavior and remains relatively constant during the 350 h measurements, deviating approximately in the {\textpm} 100 ppm range of the
initial value $f_0$ ${\approx}$ 3.5 MHz (\textbf{Figure 6}.a).

Further, the time dependent electrical conductivities of cLN-A and nsLN-A specimens are shown in Figure 6.b. As seen from the figure, the conductivity of congruent cLN-A specimen steadily decreases with the time; after 350 h of thermal treatment this decrease, however, does not exceed 3 \%, compared to the initial value $\sigma_0$ = (1.17{\textpm}0.09){\texttimes}10$^{-3}$ S/m, when the 700~°C is reached. To the contrary, the conductivity of the VTE treated sample is shown to increase by about 15\%, relative to the initial value $\sigma_0$ = (2.21{\textpm}0.18){\texttimes}10$^{-4}$ S/m. The increase is presumably caused by the increase of the electronic conduction, as discussed in Section 4.2.

    \includegraphics[width=8.059cm,height=4.953cm]{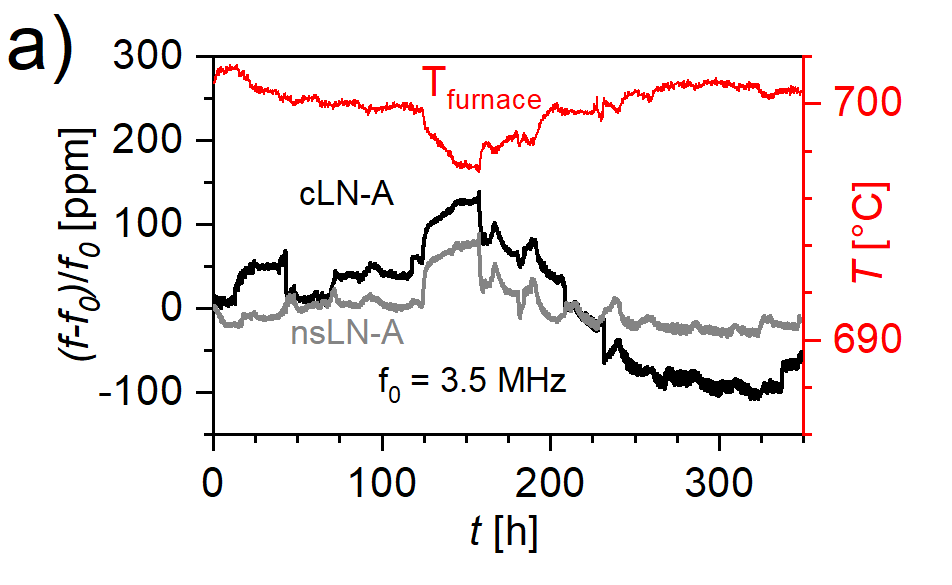} 
    \includegraphics[width=7.137cm,height=4.928cm]{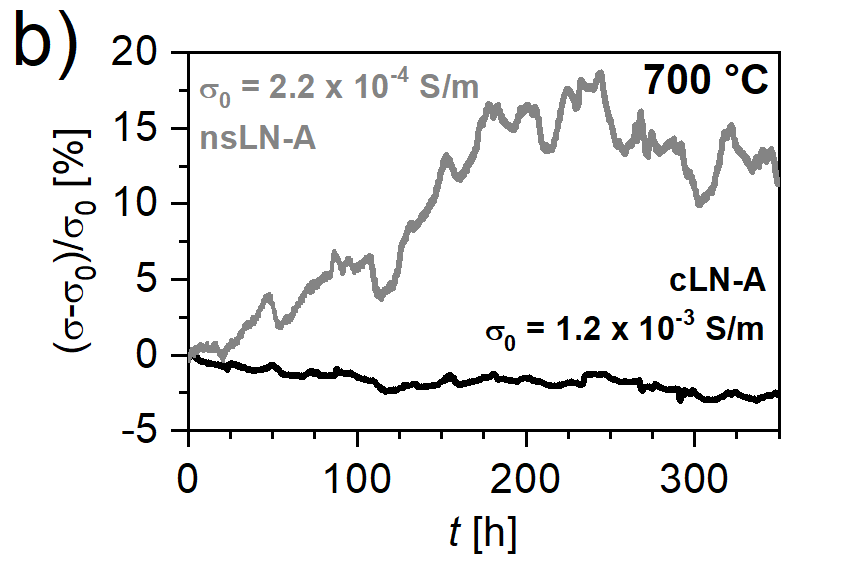} 

    \textbf{Figure 6.} Resonance frequency a) and conductivity b) deviations of cLN-A and nsLN-A samples during two weeks of annealing at 700~°C in air.

\bigskip

In addition, the absorption spectra of the samples, subjected to the long-term annealing were recorded after the treatment to detect the possible Li$_{2}$O loss from the structure. After two weeks at 700 °C, the absorption edges of both the Li-deficient and the near stoichiometric specimens were shifted towards higher wavelengths by less than 1 nm, which corresponds to about 0.1 mol\% of Li$_{2}$O loss in the case of the as-grown crystal.

\bigskip

{The changes of the electrical conductivity as a function of time, measured for 350 hours at 900~°C is shown in \textbf{Figure 7} for the cLNT45 and cLNT70 specimens.}

    {\centering  \includegraphics[width=9.188cm,height=5.655cm]{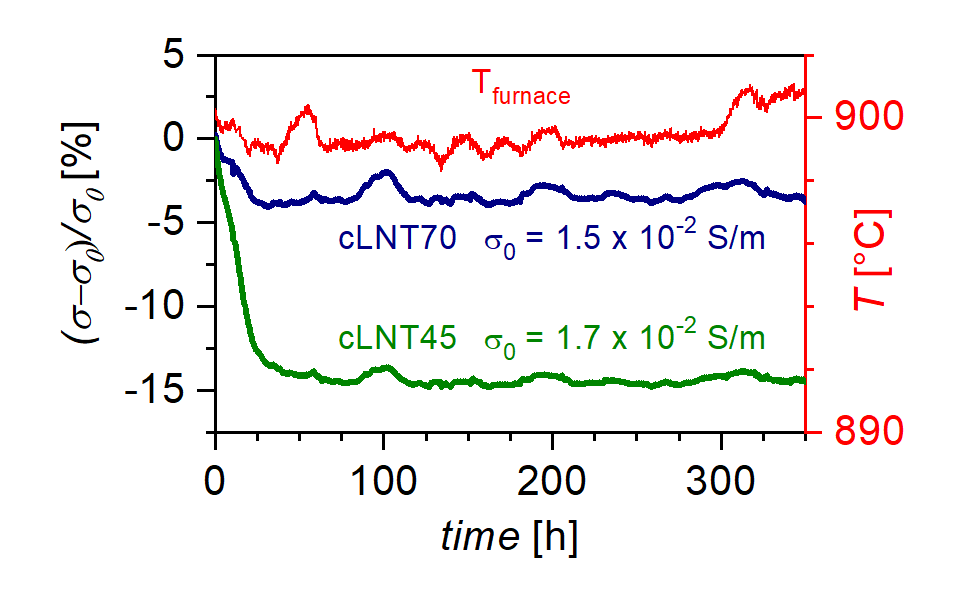} \par}
    \textbf{Figure 7.} Relative conductivity of cLNT45 and cLNT70 specimens as a function of time at 900~°C in air.

\bigskip

The initial $\sigma_0$ values, measured when the set up temperature of 900~°C was reached, were (1.68{\textpm}0.13){\texttimes}10$^{-2}$ S/m and (1.50{\textpm}0.12){\texttimes}10$^{-2}$ S/m for cLNT45 and cLNT70, respectively (see also Figure 2). As seen from Figure 7, within the first 25 hours of thermal treatment, the decrease of the conductivity by about 15\% for cLNT45 and only by about 4\% for cLNT70 is observed. Subsequently, the electrical conductivities of both specimens remain relatively constant, reaching the values of (1.45{\textpm}0.1){\texttimes}10$^{-2}$ S/m and (1.46{\textpm}0.1){\texttimes}10$^{-2}$ S/m for cLNT45 and cLNT70, respectively, after 350 hours of the thermal treatment.

\bigskip

    \section{Conclusions}

In summary, the electrical and acoustic properties of near stoichiometric Li(Nb,Ta)O$_{3}$ were studied at elevated temperatures and compared to those of the as-grown samples exhibiting Li-deficiency. The latter were subjected to the VTE treatment to achieve near stoichiometric composition. The conductivity of near stoichiometric samples was found to be lower up to 800~°C, which is attributed to the reduced concentration of lithium vacancies in these crystals. Further, it shown that up to 600--650~°C the electrical conductivity is governed by Li-ion migration. Above that temperature range, the increased contribution of different conduction mechanism is evident. Consequently, the conductivity above 650~°C is attributed to the superposition of electronic and ionic conduction. Further, the measurement of temperature-dependent $Q^{-1}$ in congruent and VTE-treated LiNbO$_{3}$ revealed rapid increase of losses at temperatures above about 650~°C, which is related to electrical conductivity and attributed to the charge carrier relaxation mechanism, superimposed on a monotonically increasing temperature-dependent background and a constant contribution. Finally, the long-term stability of LNT was examined by the measurement of electrical conductivity and resonance frequency during 350 hours of uninterrupted thermal treatment at elevated temperatures in air. It is shown for example, that the conductivity of Li-deficient specimen decreased only by about 4\% within the measuring time.

\bigskip

    \section*{Acknowledgements}

Financial support from the Deutsche Forschungsgemeinschaft (DFG, German Research Foundation)
in the framework of the research unit FOR5044 (Projects SU1261/1-1, FR1301/42-1, GA2403/7-1, FR130/40-1) is gratefully
acknowledged. Further, the authors from Clausthal University of Technology acknowledge the support of the
Energie-Forschungszentrum Niedersachsen, Goslar, Germany. M. Brützam und M. Stypa from the Leibniz-Institut für
Kristallzüchtung, Berlin, Germany are thankfully acknowledged for their technical assistance in crystal growth.

\bigskip

    {\raggedleft
    {Received: ((will be filled in by the editorial staff))}
    \par}

    {\raggedleft
    {Revised: ((will be filled in by the editorial staff))}
    \par}

    {\raggedleft
    {Published online: ((will be filled in by the editorial staff))}
    \par}

\bigskip

    \section*{References}

\begin{hangparas}{.3in}{1}
[1]\ \ O. Sánchez-Dena, C.D. Fierro-Ruiz, S.D. Villalobos-Mendoza, D.M. Carrillo Flores, J.T.~Elizalde-Galindo, R. Farías, \textit{Crystals} \textbf{2020}, 10, 973.

[2]\ \ X. Xiao, S. Liang, J. Si, Q. Xu, H. Zhang, L. Ma, C. Yang, X. Zhang, \textit{Crystals} \textbf{2023}, 13, 1233

[3]\ \ D. Xue, K. Betzler and H. Hesse, \textit{Solid State Commun.} \textbf{2000}, 115, 581.

[4]\ \ I. G. Wood, P. Daniels, R. H. Brown and A. M. Glazer, \textit{J. Phys.: Condens. Matter} \textbf{2008}, 20, 235237

[5]\ \ T. Volk, M. Wöhlecke, \textit{Lithium niobate: defects, photorefraction and ferroelectric switching}; Springer Verlag: Berlin/Heidelberg, \textbf{2008}.

[6]\ \ P.F. Bordui, R.G. Norwood, C.D. Bird, J.T. Carella, \textit{J. Appl. Phys.} \textbf{1995}, 78 4647.

[7]\ \ O.F. Schirmer, O. Thiemann, M. Wöhlecke, \textit{J.} \textit{Phys. Chem. Solids} \textbf{1991}, 52, 185.

[8]\ \ H. Donnerberg, S.M. Tomlinson, C.R.A. Catlow, O.F. Schirmer, \textit{Phys. Rev. B} \textbf{1989}, 40, 11909.

[9]\ \ D.M. Smyth, \textit{Defects and transport in LiNbO$_{3}$,\textit{Ferroelectrics}} \textbf{1983}, 50, 93.

[10]\ \ A. Vyalikh, M. Zschornak, T. Köhler, M. Nentwich, T. Weigel, J. Hanzig, R. Zaripov, E. Vavilova, S. Gemming, E.
Brendler, D.C. Meyer, \textit{Phys. Rev. Mat.} \textbf{2018}, 2, 013804.

[11]\ \ W. He, X. Gao, L. Pang, D. Wang, N. Gao, Z. Wang, \textit{J. Phys.: Condens. Matter.} \textbf{2016}, 28, 315501.

[12]\ \ P.F. Bordui, R.G. Norwood, C.D. Bird, G.D. Calvert, \textit{J. Cryst. Growth} \textbf{1991}, 113, 61.

[13]\ \ K. Lengyel, Á. Péter, L. Kovács, G. Corradi, L. Pálfalvi, J. Hebling, M. Unferdorben, G. Dravecz, I. Hajdara, Zs. Szaller, K. Polgár, \textit{Appl. Phys. Rev.} \textbf{2015}, 2, 040601.

{[14]\ \ K. Polgár, Á. Péter, L. Kovács, G. Corradi, Zs. Szaller, \textit{J. Cryst. Growth} \textbf{1997}, 177, 211.}

{[15]\ \ K. Kitamura, J.K. Yamamoto, N. Iyi, S. Kirnura, T. Hayashi, \textit{J. Cryst. Growth} \textbf{1992}, 116, 327.}

[16]\ \ M. Nakamura, S. Takekawa, Y. Furukawa, K. Kitamura, \textit{J. Cryst. }\textit{Growth} \textbf{2002}, 245, 267.

[17]\ \ P. Fielitz, O. Schneider, G. Borchardt, A. Weidenfelder, H. Fritze, J. Shi, K.D. Becker, S. Ganschow, R. Bertram, \textit{Solid State Ionics} \textbf{2011}, 189, 1.

[18]\ \ P.F. Bordui, R.G. Norwood, D.H. Jundt, M.M. Fejer, \textit{J. Appl. Phys.} \textbf{1992}, 71, 875.

[19]\ \ E.J. Samuelsen, A.P. Grande, \textit{Z. Physik B} \textbf{1976}, 24, 207.

[20]\ \ D. Damjanovic, \textit{Curr. Opin. Solid St. M.} \textbf{1998}, 3, 469.

[21]\ \ R. Fachberger, G. Bruckner, G. Knoll, R. Hauser, J. Biniasch, L. Reindl, \textit{Applicability of LiNbO}\textit{$_{3}$}\textit{, langasite and GaPO$_{4}$ in high temperature SAW sensors operating at radio frequencies}, IEEE Trans. Ultrason. Ferroelectr. Freq. Control \textbf{2004}, 51, 1427-1431.

[22]\ \ A. Bartasyte, A.M. Glazer, F. Wondre, D. Prabhakaran, P.A. Thomas, S. Huband, D.S. Keebleand, S. Margueron, \textit{Mater. Chem. Phys.} \textbf{2012}, 134, 728.

[23]\ \ L. Vasylechko, V. Sydorchuk, A. Lakhnik, Y. Suhak, D. Wlodarczyk, S. Hurskyy, U. Yakhnevych, Y. Zhydachevskyy, D. Sugak, I.I. Syvorotka, I. Solskii, O. Buryy, A. Suchocki and H. Fritze, \textit{Crystals} \textbf{2021}, 11, 755.

[24]\ \ A.M. Glazer, N. Zhang, A. Bartasyte, D.S. Keeble, S. Huband, P.A. Thomas, \textit{J. Appl. Cryst} \textbf{2010}, 43, 1305.

[25]\ \ M. Rüsing, S. Sanna, S. Neufeld, G. Berth, W.G. Schmidt, A. Zrenner, H. Yu, Y. Wang, H. Zhang, \textit{Phys. Rev. B} \textbf{2016}, 93, 184305.

[26]\ \ D. Roshchupkin, E. Emelin, O. Plotitcyna, F. Rashid, D. Irzhak, V. Karandashev, T. Orlova, \ N.~Targonskaya, S.
Sakharov, A. Mololkin, B. Redkin, H. Fritze, Y. Suhak, D. Kovalev, S. Vadilonga, L.~Ortega, W. Leitenberger, \textit{Acta Cryst. B} \textbf{2020}, 76, 1071.

[27]\ \ Y. Suhak, D. Roshchupkin, B. Redkin, A. Kabir, B. Jerliu, S. Ganschow and H. Fritze, \textit{Crystals}, \textbf{2021}, 11, 398.

[28]\ \ S. Hurskyy, U. Yakhnevych, C. Kofahl, E. Tichy-Racs, H. Schmidt, S. Ganschow, H. Fritze, Y. Suhak, \textit{Solid State Ionics},
\textbf{2023}, 399, 116285.

[29]\ \ Ch. Bäumer, C. David, A. Tunyagi, K. Betzler, H. Hesse, E. Krätzig, M. Wöhlecke, \textit{J. Appl. Phys.} \textbf{2003}, 93, 3102.

[30]\ \ L. Kovács, G. Ruschhaupt, K. Polgár, G. Corradi, M. Wöhlecke, \textit{Appl. Phys. Lett.} \textbf{1997}, 70, 2801.

[31]\ \ M. Wöhlecke, G. Corradi, K. Betzler, \textit{Appl. Phys. B} 1996, 63, 323.

[32]\ \ F. Juvalta, M. Jazbin\v{s}ek, and P. Günter, \textit{J. Opt. Soc. Am. B} \textbf{2006}, 23 (2), 276.

[33]\ \ {H. Fritze, \textit{Meas. Sci. Technol.} \textbf{2011}, 22, 12002.}

[34]\ \ A. Weidenfelder, J. Shi, P. Fielitz, G. Borchardt, K.-D. Becker, H. Fritze, \textit{Solid State Ionics,} \textbf{2012}, 225, 26.

[35]\ \ R.H. Chen, L. Chen, C. Chia, \textit{J. Phys.: Condens. Matter} \textbf{2007}, 19, 086225.

[36]\ \ B. Ruprecht, \ J. Rahn, H. Schmidt, P. Heitjans, \textit{Z. Phys. Chem.} \textbf{2012}, 226, 431.

[37]\ \ A. El-Bachiri, F. Bennani, M. Bousselamti, \textit{Spectroscopy Letters} \textbf{2014}, 47, 374.

[38]\ \ A. Krampf, M. Imlau, Y. Suhak, H. Fritze, S. Sanna, \textit{New J. Phys.} \textbf{2021}, 23, 033016.

[39]\ \ K. Lucas, S. Bouchy, P. Bélanger, R.J. Zednik, \textit{J. Appl. Phys.} \textbf{2022}, 131, 194102.

[40]\ \ A. Huanosta, A.R. West, \textit{J. Appl. Phys.} \textbf{1987}, 61, 5386.

[41]\ \ D.C. Sinclair, A.R. West, \textit{Phys Rev B} \textbf{1989}, 39, 13486.

[42]\ \ D. Ming, J.M. Reau, J. Ravez, J. Gitae, P. Hagenmuller, \textit{J. Solid State Chem.} \textbf{1995}, 116, 185.

[43]\ \ U. Yakhnevych, C. Kofahl, S. Hurskyy, S. Ganschow, Y. Suhak, H. Schmidt, H.Fritze, \textit{Solid State Ionics} \textbf{2023}, 392, 116147.

[44]\ \ U. Yakhnevych, F. El Azzouzi, F. Bernhardt, C. Kofahl, Y. Suhak, S. Sanna, K.-D. Becker, H. Schmidt, S. Ganschow, H. Fritze, \textit{ Oxygen partial pressure and temperature dependent electrical conductivity of Lithium Niobate-Tantalate solid solutions}, submitted to Solid State Ionics \textbf{2023}

[45]\ \ J. Rahn, P. Heitjans, H. Schmidt, \textit{J. Phys. Chem. C} \textbf{2015}, 119, 15557.

[46]\ \ J.R. {\textup{Carruthers}}, G.E. Peterson, M. Grasso, \textit{J. Appl. Phys.} \textbf{1971}, 42, 1846.

[47]\ \ M. Katz, P. Blau, B. Shulga, \textit{Proc. of SPIE} \textit{Nonlinear Frequency Generation and Conversion:
Materials, Devices, and Applications VII} \textbf{ 2008}, 6875, 687504.

[48]\ \ F. El Azzouzi, U. Yakhnevych, Y. Suhak, D. Klimm, L. Verhoff, N. Schäfer, H. Schmidt, S. Ganschow, S. Sanna, K.-D. Becker, H. Fritze, \textit{Phase Transformation in Lithium Niobate-Lithium Tantalate ({LiNb$_{1-x}$Ta$_{x}$O$_{3}$}) Solid Solutions}, submitted to Physica Status Solidi A \textbf{2023}

[49]\ \ R.D. Shannon, \textit{Acta Crystallogr. A} \textbf{1976}, 32, 751.

[50]\ \ P. Fielitz, G. Borchardt, S. Ganschow, R. Bertram, R. Jackson, H. Fritze, K.-D. Becker, \textit{Solid State Ionics} \textbf{2014}, 259, 14.

[51]\ \ P. Fielitz, G. Borchardt, R.A. De Souza, M. Martin, M. Masoud, P. Heitjans, \textit{Solid State Sci.} \textbf{2008}, 10, 746.

[52]\ \ C. Kofahl, L. Dörrer, B. Muscutt, S. Sanna, S. Hurskyy, U. Yakhnevych, Y. Suhak, H.~Fritze, S. Ganschow, H. Schmidt, \textit{Phys. Rev. Mater.} \textbf{2023}, 7, 033403.

[53]\ \ L. Dörrer, P. Tuchel, E. Hüger, R. Heller, H. Schmidt, \textit{J. Appl. Phys.} \textbf{2021}, 129, 135105.

[54]\ \ K. Ishibashi, Y. Okuyama, N. Fukatsu, \textit{J. Jpn. I. Met. Mater.} \textbf{2011}, 75, 229.

[55]\ \ D. Smyth, \textit{Ferroelectrics} \textbf{1983}, 50, 419.

[56]\ \ G. Bergmann, \textit{Solid State Commun.} \textbf{1968}, 6, 77.

[57]\ \ P. Jorgensen, R. Bartlett, \textit{J. Phys. Chem. Solids} \textbf{1969}, 30, 2639.

[58]\ \ O. F. Schirmer, M. Imlau, C. Merschjann, B. Schoke, \textit{J. Phys.: Condens. Matter} \textbf{2009}, 21, 123201.

[59]\ \ A. Ballato, \textit{IEEE Trans. Instr. Meas.}\textbf{1978}, 27, 59.

[60]\ \ J.J. Martin, A.R. Lopez, \textit{High-temperature acoustic loss in. AT-cut, BT-cut and SC-cut quartz resonators}, Proc. IEEE/PDA Int. Freq. Control Symp. (2001) 316-323.

[61]\ \ W.L. Johnson, S.A. Kim, S. Uda and C.F. Rivenbark, \textit{J. Appl. Phys} \textbf{2011}, 110, 123528.

[62]\ \ C. Hirschle, J. Schreuer \textit{, IEEE Trans. Ultrason., Ferroelect., Freq. Control} \textbf{2018}, 65 1250.

[63]\ \ Yu. Suhak, M. Schulz, W. L. Johnson, A. Sotnikov, H. Schmidt, H. Fritze, \textit{Solid State Ionics} \textbf{2018}, 37, 221.

[64]\ \ I. Kogut, C. Hartmann, I. Gamov, Y. Suhak, M. Schulz, S. Schröder, J. Wollweber, A. Dittmar, K. Irmscher, T. Straubinger, M. Bickermann, H. Fritze, \textit{Solid State Ionics} \textbf{2019}, 343, 115072.

[65]\ \ A. R. Hutson, D. L. White, \textit{J. Appl. Phys.} \textbf{1962}, 33, 40.

[66]\ \ T. Ikeda, \textit{Fundamentals of piezoelectricity}; Oxford University Press: Oxford, UK, \textbf{1990}.

[67]\ \ H. De Castilla, P. Belanger, R.J. Zednik, \textit{J. Appl. Phys.} \textbf{2017}, 122, 244103.
\end{hangparas}

\end{document}